\documentstyle[12pt,epsf]{article}
\def\SOD{\sigma_{od}}\def\sod{\ifmmode\SOD\else$\SOD$\fi}
\def\SOO{\sigma_{oo}}\def\soo{\ifmmode\SOO\else$\SOO$\fi}

\newcommand{\en}[1]{\mbox{$E^{#1}$}}
\newcommand{\beq}{\begin{equation}} \newcommand{\eeq}{\end{equation}}
\newcommand{\bfg}{\begin{figure}[htb]} \newcommand{\efg}{\end{figure}}

\def\etal{{\sl et al.\/}}
\def\CMP{{\sl Comm.\ Math.\ Phys.\/}}    \def\JSP{{\sl J.\ Stat.\ Phys.\/}}
\def\IJMP{{\sl Int.\ J.\ Mod.\ Phys.\/}} \def\PL{{\sl Phys.\ Lett.\/}}
             \def\PRL{{\sl Phys.\ Rev.\ Lett.\/}}
\def\NP{{\sl Nucl.\ Phys.\/}}            \def\ZP{{\sl Z.\ Phys.\/}}
\begin{document}
\thispagestyle{empty}     
\hbox{} \mbox{}\hspace{1.0cm}October 1993 \hspace{8.0cm} HLRZ 63/93\\
\begin{center} \vspace*{1.8cm}
{\large INTERFACE TENSIONS \\AND PERFECT WETTING\\
        IN THE TWO-DIMENSIONAL\\ SEVEN-STATE POTTS MODEL}\\
\vspace*{1.0cm}
{\large B.~Grossmann$^1$ and Sourendu Gupta$^{1,2}$\\ }
{\normalsize\vspace*{1.0cm}
$\mbox{}^1$ {HLRZ, c/o Research Center Juelich, D-52425 Juelich, Germany}\\
$\mbox{}^2$ {TIFR, Homi Bhabha Road, Bombay 400005, India.}\\}
\vspace*{3.0cm} {\large \bf Abstract} \end{center}
\setlength{\baselineskip}{1.3\baselineskip}

We present a numerical determination of the order-disorder interface
tension, \sod, for the two-dimensional seven-state Potts model. We
find $\sod=0.0114\pm0.0012$, in good agreement with expectations
based on the conjecture of perfect wetting. We take
into account systematic effects on the technique of our choice:
the histogram method. Our measurements are performed on rectangular
lattices, so that the histograms contain identifiable plateaus.
The lattice sizes are chosen to be large compared to the physical
correlation length. Capillary wave corrections are applied to our
measurements on finite systems.
\newpage\setcounter{page}1

Phase transitions of interest to cosmology and colliders may well turn
out to be of first order. For this reason there has been increased
interest in first-order transitions. Among the parameters governing the
dynamics of such transitions are the interface tensions. These are
properties of the equilibrium system. There have been attempts to
measure these on the gauge theories of relevance to high-energy physics.
Recently, however, the control of systematic biases in these
measurements have been questioned. Such uncertainties call for detailed
tests of the methodology in models which are more tractable than gauge
theories.

The two-dimensional Potts models are used very often for such purposes.
They are also interesting because of connections with exactly solved
two-dimensional models in statistical mechanics. They are defined by
the partition function
\beq
  Z\;=\;{\rm Tr}\exp(-\beta H), \qquad{\rm where}\qquad
  H\;=\;-\sum_{\langle ij\rangle}\delta_{\sigma_i\sigma_j}.
\label{eq:potts}\eeq
The spins $\sigma_i$ at each site $i$ of a lattice can take $q$
different values, and the notation $\langle ij\rangle$ denotes nearest
neighbour pairs. There is an order-disorder phase transition in the
models for all $q$, with a singlet magnetisation as an order-parameter.
This transition occurs at a coupling $\beta_c=\log(\sqrt{q}+1)$. For
$q>4$ the transition is of first-order.

The coexisting phases can be characterised by the internal energy
densities
\beq
E\;=\;\frac{1}{V}\langle H\rangle,
\label{eq:ene}\eeq
where the angular brackets denote thermal expectation values and $V$ is
the volume of the lattice. In the ordered and disordered phases at
$\beta_c$, the values of the internal energy densities will be denoted
by \en o and \en d respectively. For the Potts models these values are
known exactly \cite{baxter}.
Recall that as $\beta\to\beta_c^+$ we obtain a pure ordered phase with
$E=\en o$. As the system is heated, its internal energy increases at
constant $\beta$. This is achieved by creating larger and larger volume
fractions of the disordered phase. Coexistence of the two phases implies
the creation of interfaces. Finally, when $E=\en d$, a purely disordered
phase is obtained and with additional heating $E$ increases with a
decrease of $\beta$. In the coexistence region, $\en o <E<\en d$ at
$\beta_c$, there is a non-extensive part of the free energy. This part
can be used to define the interface tension.

A recent exact computation of the correlation length at $\beta_c$ in
all Potts models with $q>4$ \cite{buf92} has been identified with the
correlation length in the disordered phase, $\xi_d$, on the basis of
a large-$q$ expansion \cite{bor92}. A duality argument \cite{laa87}
then gives the interface tension between two different ordered phases,
\soo, as
\beq
\soo\;=\;1/\xi_d.
\label{eq:soo}\eeq
Now making the assumption of perfect wetting,
\beq
\soo\;=\;2\sod,
\label{eq:pw}\eeq
one obtains the prediction
\beq
\sod\;=\;1/2\xi_d.
\label{eq:pred}\eeq
Since $\xi_d$ is exactly known, a measurement of \sod{} would check the
relation in eq.~(\ref{eq:pred}) and thus test the conjecture of perfect
wetting in eq.~(\ref{eq:pw}). This assumption is for large $q$ by the
proof of the opposite inequalities $\soo\ge2\sod$ and $\soo\le2\sod$
\cite{mixed}. A recent measurement at $q=15$ \cite{gup93} has used the
argument outlined here to verify that perfect wetting holds for that model.

The status of the model with $q=7$ is interesting. With the assumption
of perfect wetting, the exactly known value of $\xi_d$ would predict
$\sod=0.010396\cdots$. A first measurement using the so-called integral
method gave $\sod=0.0943(6)$ \cite{pot89}. A value in agreement,
$\sod\approx0.10$ was obtained \cite{kaj89} by a different technique.
The surprisingly large violation of perfect wetting led to criticism
of these two techniques for the measurement of interface tensions.
Two measurements using the histogram technique were then performed.
The first gave $\sod=0.0121(5)$ \cite{jan92} and the second
$\sod=0.01174(19)$ \cite{rum92}. Although they are in agreement with
each other, they are in violation of perfect wetting by about seven
standard deviations. The lattice sizes used in both these studies were
not much larger than $\xi_d$, and hence the results may require
correction. This was also noticed in \cite{bha93}. In this letter we
report a study of the $q=7$ model by the histogram technique
correcting for all sources of systematic errors.

Consider a finite lattice of size $L\times L_z$ ($L_z>L$), with
periodic boundary conditions imposed on the configuations. The histogram
of the probability distribution of $E$, $P_L(E)dE$, was first used for
the extraction of an interface tension in \cite{binder}. $P_L(E)$ has
peaks at \en o and \en d, shifted through finite size effects. In an
interval of $E$ bounded by these two values, there is a minimum,
$P_L^{min}$, and a nearly flat region around it. This region is due to
two interfaces spanning the smaller direction, $L$, and seperating an
ordered from a disordered phase. The histogram method consists of a
model for $P_L^{min}$ and the nearly flat plateau contained in the
interval $\en o<E<\en d$. Such a model contains as a parameter the
interface tension of interest, which is then extracted by comparison
with data.

In the region of interest, $P_L(E)$ is independent of $E$ as long as the
free energy is independent of the relative positions of the two
interfaces. Fluctuations of the interface shapes are described by the
capillary wave model~\cite{pri92}. For non-interacting interfaces, one
obtains \cite{bun92}
\beq
P_L^{min}\;=\;{L_z^2\over L}\exp\left( -2\sod L\right).
\label{eq:capwav}\eeq
This contribution takes into account interfaces perpendicular to the
$z$-direction. For $L_z>L$, interfaces parallel to the $z$-direction
give small corrections to this result. For the lattices we have chosen
to work with, these corrections are negligible.
In \cite{wie92} it was shown that the formula in eq.~(\ref{eq:capwav})
holds for two-dimensional systems even when the global condition that
$E$ lies between the two pure-phase energies is imposed.

Whenever all the above conditions are fulfilled, one can extract the
interface tension from either of the two quantities defined below. The
pre-exponential factors in eq.~(\ref{eq:capwav}) cancel in
\beq
F_L^1\;=\;\frac{1}{2L}\ln\frac{\overline{P_L}^{max}}{P_L^{min}}
            -\frac{3}{4}\frac{\ln L}{L}+\frac{3}{4}\frac{\ln L_z}{L}.
\label{eq:fl1}\eeq
where $\overline{P_L}^{max}$ is the geometric mean of the maxima of
$P_L(E)$ corresponding to the ordered and disordered phases. Taking into
account capillary wave fluctuations of the interfaces, we obtain the
finite size scaling formula
\beq
F_L^1\;=\;\sod+\frac{b_1}{L} \qquad{\rm for\ } L_z>L\to\infty.
\label{eq:fl1fss}\eeq
As a cross-check one can also measure a second quantity from the
histograms,
\beq
F_L^2\;=\;-\frac{1}{2L}\ln P_L^{min}
           -\frac{1}{2}\frac{\ln L}{L}+\frac{\ln L_z}{L}.
\label{eq:fl2}\eeq
Capillary wave fluctuations of the two interfaces also yield the finite
size scaling formula for this quantity---
\beq
F_L^2\;=\;\sod+\frac{b_2}{L} \qquad{\rm for\ } L_z>L\to\infty.
\label{eq:fl2fss}\eeq
We shall use both the quantities, $F_L^1$ and $F_L^2$, along with their
finite size scaling formul\ae{} to estimate the interface tension.

\begin{table}[hbt]\begin{center}\begin{math}
\begin{array}{|c|c|c|c|}\hline
L\times L_z&\beta&N/10^6\\\hline
60\times120&1.293&2.1\\
80\times160&1.293&4.5\\
100\times200&1.2931&5.0\\
140\times280&1.293432&20.0\\
            &1.293562&12.0\\ \hline
\end{array}\end{math}\end{center}
\caption{The lattice sizes, couplings and the number of sweeps, $N$,
   used in this study.}
\label{tab:simulations}\medskip\noindent\end{table}

Care is required to ensure that the relations above may be applied to
the data obtained. A sufficient check is to confirm that $P_L(E)$
indeed has a flat plateau at its minimum. In \cite{gro93} it was found
that the condition $L_z>L$ was necessary to obtain non-interacting
interfaces. Previous studies of this model had used only square lattices.
In this study we consistently use the value $L_z=2L$. It has turned out
 that the lattices need not be elongated further for a
flat plateau in the histogram to develop. A second necessary condition is
that the system break up into several domains. This requires $L_z\gg\xi$.
All our lattices satisfy this condition, whereas earlier studies were
performed on much smaller lattices. Finally, we note that the capillary
wave model of interface fluctuations has been applied successfully in other
contexts, and is bound to be an improvement over the neglect of interface
fluctuations.

\bfg\centerline{\epsfysize=11cm\epsfbox{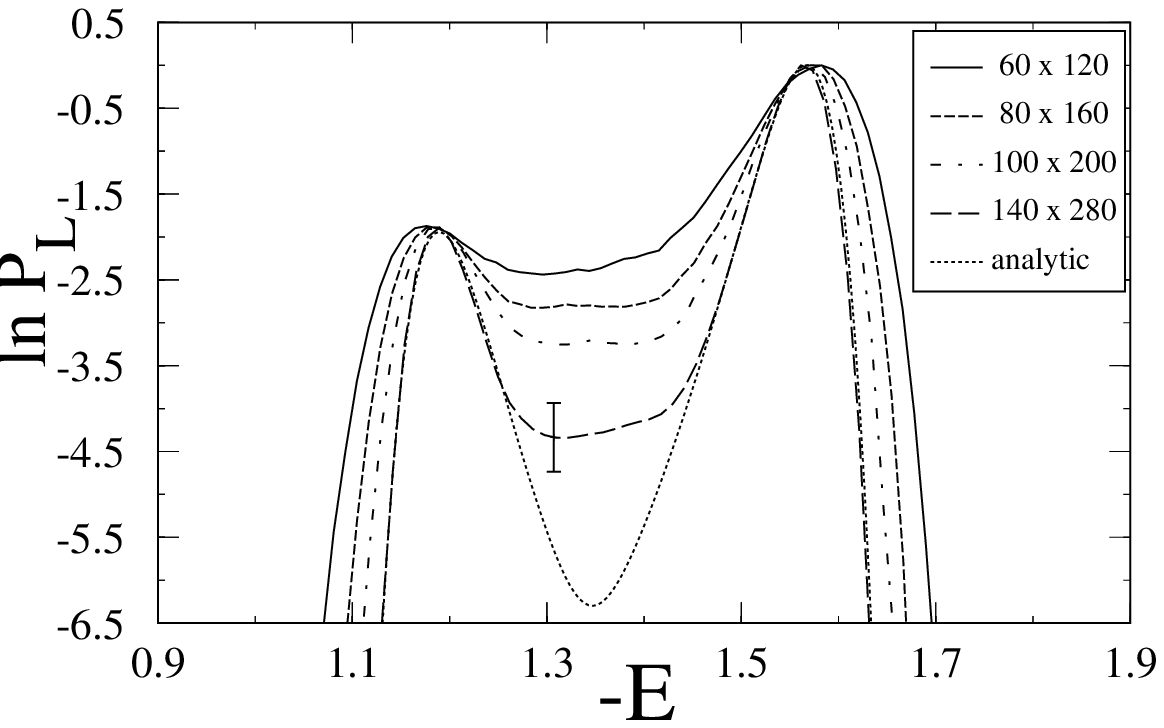}}
\caption{{Measured probability distributions for the energy density $E$.
   The dotted line shows the result of a computation following
   \protect\cite{bha93} for the largest lattice.}}
\label{fig:dist}\efg

The simulations were performed with $L\times2L$ lattices. Since the
longest physical correlation length at $\beta_c$ is
$\xi_d=48.09\cdots$, the minimum value of $L$ used was 60. For $L\le100$
we used a cluster update. For larger values of $L$ the performance of
this algorithm was not satisfactory. Hence we used a multicanonical
version of a two hit Metropolis algorithm in the form specified in
\cite{algo}. The energy was measured every tenth sweep. Between 1000 and
5000 initial configurations were discarded for thermalisation. The
couplings were chosen so that the canonical weights were roughly equal
for the two phases. The second run on the $L=140$ lattice was an
exception, since this was performed at $\beta_c$. The data was
reweighted to the canonical distribution at $\beta_c$ for each lattice.
A cross-check on the statistics of our runs was that the weights of the
two peaks at $\beta_c$ are in the ratio 1:7. For the $L=140$ lattice,
the results presented later average over the two runs.

We show the histograms obtained in our computations in Figure
\ref{fig:dist}. We would like to emphasise that flat plateaus
are obtained over an interval between the two peaks for $L\ge80$.
In a criticism of the histogram method as applied to the ten-state
Potts model on square lattices \cite{bha93}, it was argued that
$P_L(E)$ could be reproduced by an analytic computation which neglects
all interfaces which span the lattice. This is not true for the lattice
sizes and shapes we choose to work with. In Figure \ref{fig:dist} we
show our histograms and a comparison with a computation following the
methods of \cite{bha93}. The analytic computation strongly
underestimates the probability in the intermediate region of $E$.
We also find that the discrepancy increases with increasing $L$. This is
evidence that interface contributions are important in this region of $E$.

\bfg\centerline{\epsfysize=11cm\epsfbox{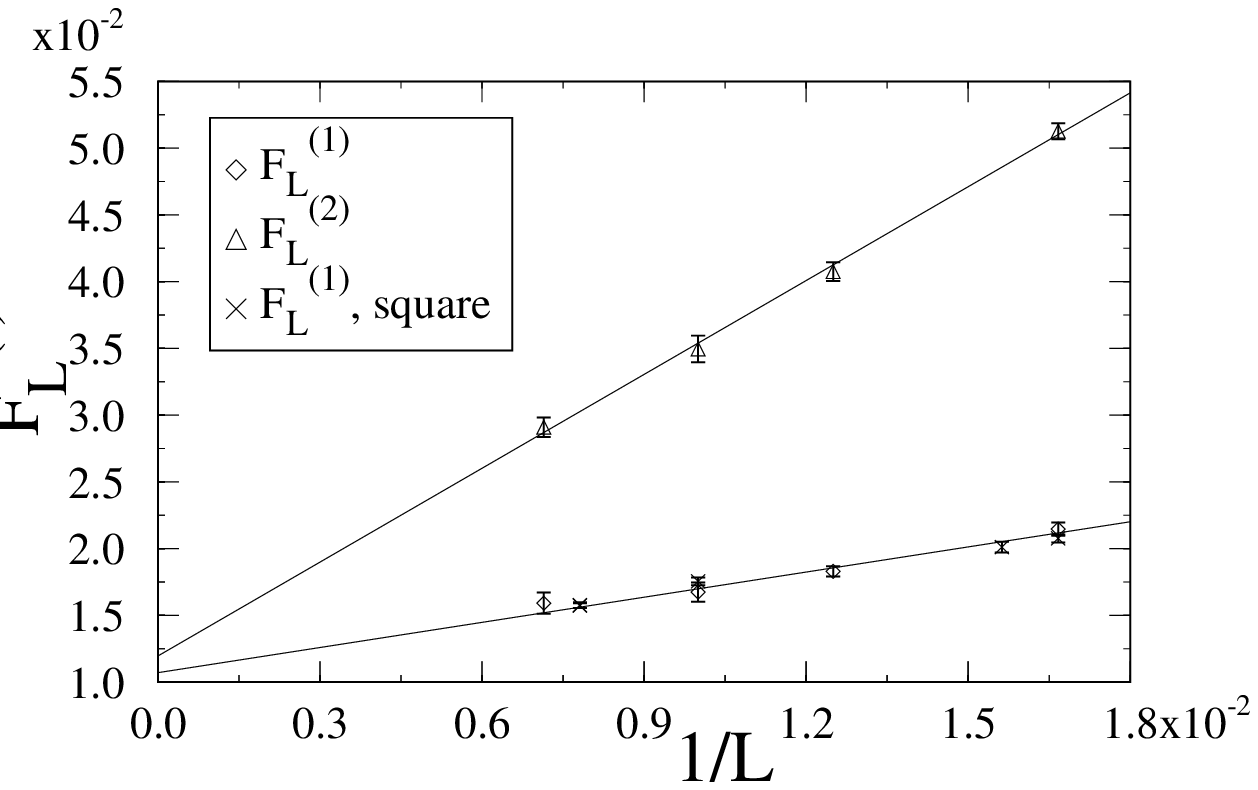}}
\caption{{Results for $F_L^i$ together with linear fits. The crosses are
   results for square lattices from \protect\cite{jan92,rum92}.}}
\label{fig:fl}\efg

We extract $F_L^1$ and $F_L^2$ from the histograms. A jack-knife
procedure is used for the estimates of the averages and errors. The
results quoted were obtained using between 6 and 24 jack-knife blocks.
We checked that the results were stable under change in the number of
blocks. A finite size scaling analysis of each of these measurements is
performed by fits to the data. Such an analysis gives
\beq
\sod\;=\;\cases{
      0.0107 (12) & \qquad from\ $F_L^1$ ($\chi^2=1.73/2$\ dof),\cr
      0.0120 (12) & \qquad from\ $F_L^2$ ($\chi^2=1.18/2$\ dof).}
\label{eq:result}\eeq
An overall estimate, obtained by averaging these two values, is
\beq
\sod\;=\;0.0114\pm0.0012.
\label{eq:final}\eeq
The data and fits leading to this result are displayed in Figure
\ref{fig:fl}.

It is possible to perform an interesting exercise with the data on
$F_L^1$ obtained in \cite{jan92,rum92}. Recall that this data was
obtained through simulations in square lattices where the existence of
the plateau due to interfaces was not clear. Nevertheless, a rough
correction may be performed if one argues that the interfaces could be
either parallel or perpendicular to the $z$-direction, and hence the net
result would be to multiply $P_L^{min}$ in eq.~(\ref{eq:capwav}) by a
factor of two. Applying this correction, one obtains from the square
lattice results a value for \sod{} surprisingly close to our result
quoted in eq.~(\ref{eq:final}). Note, however, that in the absence of a
clear plateau around $P_L^{min}$ this result is accidental.

In summary, we have performed a measurement of the order-disorder
interface tension, \sod, in the two-dimensional seven-state Potts model
and found a value in close agreement with the result expected on the
basis of perfect wetting. The measurement technique used was the
so-called histogram method. We have demonstrated that using rectangular
lattices of shape $L\times2L$ where $L\gg\xi$, one may overcome previous
objections to this method and clearly obtain a plateau at the minimum of
the histogram. This, and the use of capillary wave corrections to the
interface shape are sufficient condition for a reliable application of
this technique to the measurement of interface tensions.

We thank Andr\'e Morel for computing the probability distributions for
the seven-state Potts model for the lattice sizes used here, following
the methods of \cite{bha93}. We would also like to thank Alain Billoire,
Robert Lacaze, and Andr\'e Morel for interesting discussions. One of us
(SG) would like to acknowledge the hospitality at TFT, Helsinki while
this paper was being written. The computations were performed on the
Cray Y/MP at HLRZ, J\"ulich.

\vfil\eject


\begin{thebibliography}{10}
\bibitem{baxter}
   R.~J.~Baxter, {\sl Exactly Solved Models in Statistical Mechanics},
   Academic Press, London, 1982.
\bibitem{buf92}
   E.~Buffenoir and S.~Wallon, {\sl J.\ Phys.\/} A 26 (1993) 3045.
\bibitem{bor92}
   C.~Borgs and W.~Janke, {\sl J.\ Phys.\ (France)\/} I 2 (1992) 2011.
\bibitem{laa87}
   L.~Laanait, \PL~A 124 (1987) 480.
\bibitem{mixed}
   J.~de Coninck \etal, \JSP~52 (1988) 45;\\
   R.~Schonmann, \JSP~52 (1988) 61;\\
   A.~Messager \etal, \CMP~140 (1991) 275.
\bibitem{gup93}
   Sourendu Gupta, HLRZ preprints HLRZ-22/93 and HLRZ-65/93.
\bibitem{pot89}
   J.~Potvin and C.~Rebbi, \PRL~62 (1989) 3062.
\bibitem{kaj89}
   K.~Kajantie, L.~K\"arkk\"ainen, and K.~Rummukainen,
   \PL~B 223 (1989) 213.
\bibitem{jan92}
   W.~Janke, \IJMP~C 3 (1992) 1137.
\bibitem{rum92}
   K.~Rummukainen,  \NP~B ({\sl Proc.\ Suppl.\/}) 30 (1993) 273.
\bibitem{bha93}
   T.~Bhattacharya, R.~Lacaze, and A.~Morel,
   Saclay preprint SPhT-93-022, 1993.
\bibitem{binder}
   K.~Binder, \ZP~B 43 (1981) 119.  
\bibitem{pri92}
   V.~Privman, \IJMP~C 3 (1992) 857.
\bibitem{bun92}
   B.~Bunk, \IJMP~C 3 (1992) 889.
\bibitem{wie92}
   U.~J.~Wiese, preprint BUTP-92/37, 1992, to appear in \JSP.
\bibitem{gro93}
   B.~Grossmann and M.~L.~Laursen, HLRZ preprint, HLRZ-93-7,
   (to appear in \NP~B).
\bibitem{algo}
   B.~Grossmann, M.~L.~Laursen, T.~Trappenberg, and U.-J.~Wiese,
   \PL~B 293 (1992) 175.
\end{thebibliography}
\end{document}